\documentstyle[multicol,aps,prl,floats]{revtex}

\begin{document}
\title{Propagation of a Huygens front through turbulent medium}
\author{M. Chertkov{$^a$} and V. Yakhot{$^b$}}
\address{Physics Department$^{a}$ and Programm in Applied and Computational
Mathematics$^{b}$,\\
Princeton University, Princeton, NJ 08544, USA}
\date{September 25, 1997}
\maketitle

\begin{abstract}
The dynamics of a thin Huygens front propagating through turbulent medium is
considered. A rigorous asymptotic expression for the effective velocity $%
v_{F}$ proportional to the front area is derived. The small-scale
fluctuations of the front position are shown to be strongly intermittent.
This intermittency plays a crucial role in establishing a steady state
magnitude of the front velocity. The results are compared with experimental
data.
\end{abstract}

\draft

\begin{multicols}{2}

The problem of propagation of a thin passive front (of flame,
phase-transition, etc.) through turbulent flow has attracted a lot of
attention since early forties \cite{40Dam,47Sch} when it was realized that
velocity fluctuations tend to generate strongly convoluted front,
thus dramatically increasing their area. In premixed combustion processes
the flame front area is directly related to the speed of the front
propagation. The first expression for the flame front velocity $v_{F}\propto
u_{rms}$, valid in the limit $u_{rms}\gg u_{0}$, where $u_{rms}$ is the
root-mean-square velocity of the integral eddy and $u_{0}$ is the laminar
flame speed depending on the details of chemical kinetics, was proposed by
Schelkin \cite{47Sch} (see \cite{85Wil,87Pop} for modern reviews on the
theory of turbulent combustion). The problem is also important for
description of light propagation in the media with fluctuating dielectric
constant, shock wave fronts etc. \cite{93CH}. The renormalization group
approach \cite{88Yak}, which included some theoretically unjustified steps,
yielded the expression, differing from the Schelkin result by a logarithmic
factor, which agreed with experimental data \cite{92SRBY,95RHR,96SJR} in a
wide range of parameter variation. Still, despite substantial activity, no
rigorous derivation of the front speed appeared and the question of the
dependence of $v_{F}$ on both $u_{0}$ and the Reynolds number $Re$ remained
open. It is shown in this work that recent advances in the theory of a
passive scalar, advected by turbulence, enable one to accurately account for
the small-scale intermittency of a scalar field, crucial for description of
the front fluctuations and derivation of the effective velocity $v_{F}$.

We consider a problem of propagation of a passive front though turbulent
flow. The front can be described by the equation for a passive scalar \cite
{85Wil,88KAW} (so called ''G''-equation) 
\begin{equation}
\partial _{t}G+({\bf v}\partial _{{\bf r}})G=u_{0}|\partial _{{\bf r}}G|,
\label{fpe}
\end{equation}
where $G(t,{\bf r})$ is a scalar field whose level surface, say $G=0$,
represents the thin front position. Statistics of turbulent velocity ${\bf v}
$ is supposed to be known. The equation (\ref{fpe}) with ${\bf v}=0$
describes a front propagating with the constant speed $u_{0}$ (laminar front
speed) normally to the local orientation of the front. For example, if the
front at time $t=0$, is defined at the $x-y$ plane, it will propagate with a
constant speed $u_{0}$ and constant area $S_{0}=const$ in the $z$-
direction. The role of the random field ${\bf v}$ is in generation of a
strongly convoluted (`` wrinkled'') front with a substantially increased
area $S_{T}>S_{0}$. When ${\bf v}=0$ the mass of the reagents (fuel)
consumed per unit time is: $dm/dt=u_{0}S_{0}=const$ does not change in time.
In general, $dm/dt=u_{0}S_{T}\equiv v_{F}S_{0}$, where $S_{T}$ is the area
of the wrinkled front. This gives a definition of a turbulent or effective
front velocity, $v_{F}=u_{0}S_{T}/S_{0}$. Assuming that a steady (both $%
v_{F} $ and $S_{T}$ do not depend on time after a long evolution) regime is
realized, we introduce a new variable, $G(t,{\bf r})\equiv v_{F}t-z+h(t,{\bf %
r})$. Then, (\ref{fpe}) reads: 
\begin{equation}
\partial _{t}h+({\bf v}\partial _{{\bf r}})h=v_{z}+u_{0}\sqrt{\left(
\partial _{{\bf r}}h\right) ^{2}+1-2\partial _{z}h}-v_{F}.  \label{fpe1}
\end{equation}
In the moving frame the fluctuations of the front position $h$ are assumed
to be in a statistically steady state, that fixes the value of the front
speed $v_{F}$, which in the turbulent regime discussed ( $v_{F}\gg u_{0}$,
when $|\partial _{r}h|^{2}\gg |\partial _{z}h|\gg 1$) is given by 
\begin{equation}
v_{F}=u_{0}\langle |\partial _{{\bf r}}h|\rangle \sim u_{0}\left\langle
|\delta h(r_{0})|\right\rangle /r_{0}.  \label{first}
\end{equation}

\noindent Here, $|\delta h(r_{0})|$ is a magnitude of the velocity
difference at the scale $r_{0}$ where the "chemical" ($u_0$-dependent) and
advective contributions to (\ref{fpe1}) balance each other. Averaging over
the turbulent velocity is assumed in (\ref{first}). Since $u_{rms}\gg u_{0}$%
, $r_{0}\ll L$, where $L$ is the scale of the turbulence source. This means
that the scalar, injected at the scale $L$, is dissipated at the propagating
front as a result of generation of very sharp cusps of the radius $r_{0}<<L$%
. Formation of such cusps was observed in numerical simulations \cite
{88ASY,88OS}. Derivation of characteristic width $r_{0}$, magnitude $|\delta
h(r_{0})|$ of these cusps and, as a result, $v_{F}$, is the goal of the
theory.

The presence of the two characteristic scales $\eta $ and $r_{0}$ defines
three possible flame regimes:

\begin{equation}
\frame{A:\hspace{0.1in}L$\gtrsim \eta \gg $r$_{0}$},\hspace{0.1in}\frame{B:%
\hspace{0.1in}L$\gg $r$_{0}\gg \eta $},\hspace{0.1in}\frame{C:\hspace{0.1in}L%
$\gg \eta \gg $r$_{0}$.}  \label{C}
\end{equation}
Generation of the small-scale scalar fluctuations (direct cascade) \cite
{49Obu,51Cor} takes place in both inertial-convective, $L>r>\eta $ and the
dissipative-convective, $\eta >r>r_{0}$ (which is valid in the $A,C$ cases
but not $B$) intervals. Thus, the problem is naturally divided into two:
First, we need to describe scalar (height of the flame brush) correlations
in both convective ranges $L>r\gg r_{0}$. Once the solutions in the
convective intervals are found, we will be ready to resolve the second and
principal part of the problem: to calculate the value of the dissipative
scale $r_{0}$ and, matching dissipative and convective intervals, find
turbulent speed of the front, $v_{F}$. Therefore, naturally, we are starting
from the first task, considering all the regimes ( $A-C$) one after another.

\underline{A.} Batchelor (a pure viscous - convective) regime. The case $A$
corresponds to a well- studied situation, first discussed by Batchelor \cite
{59Bat} and developed further in \cite{67Kra,94SS,95CFKLa}. Without loss of
generality and following \cite{95CFKLa} we will consider the velocity
difference to be Gaussian in the case 
\begin{equation}
\left\langle \delta v_{r}^{\alpha }(t)\delta v_{r}^{\beta }(t)\right\rangle =%
\frac{D}{\tau }\left[ 2\delta ^{\alpha \beta }r^{2}-r^{\alpha }r^{\beta
}\right] \exp \left[ -\frac{t}{\tau }\right] ,  \label{modelA}
\end{equation}
where $\tau $ is turn-over time of the integral ( $L$-size) eddy. The pair
correlation function of the scalar obeys the famous logarithmic law in the
convective interval of scales, $L\gg r\gg r_{0}$, \cite{67Kra,94SS,95CFKLa} 
\begin{equation}
\langle h_{1}h_{2}\rangle =DL^{2}\ln \left[ L/r_{12}\right] /\lambda ,
\label{log1}
\end{equation}
where $DL^{2}$ in the nominator of the logarithmic prefactor describes the
fluctuation of the ''source'' function $v_{z}$ while $\lambda $ stands for
the Lyapunov exponent corresponding to the rate of Lagrangian stretching.
Correlation between the source and convective terms in (\ref{fpe1}) does not
contribute to (\ref{log1}). Accounting for these correlations slightly
modifies the higher-order moments generating subleading contributions and a
mere renormalization of bare coefficients. The Lyapunov exponent $\lambda $
as a function of $\tau ,$ $D$ was found for the two-dimensional version of
the model (\ref{modelA}) in \cite{95CFKLa}. Asymptotic of the large and
small $\tau $ were described explicitly in \cite{95CFKLa}. Generally, the
problem of finding $\lambda $ was reduced in \cite{95CFKLa} to a
well-defined auxiliary quantum mechanics which was easy to solve
numerically. An interpolation formula for $\lambda ,$ fitting well all the
known asymptotic is: 
\begin{equation}
\lambda =\sqrt{D/\tau }\tanh \left[ \sqrt{D\tau }\right] .  \label{lam}
\end{equation}
This formula, valid in the space of arbitrary dimensionality $d>2$, holds up
to ${\cal O}(d)$ corrections.  Statistics of the scalar fluctuations in the
convective interval is shown to be Gaussian \cite{95CFKLa}. Therefore, the
typical fluctuation of the height $h$ at $r_{0}$, estimated by the second
moment (\ref{log1}) is
\begin{equation}
|\delta h_{r_{0}}|\sim L\sqrt{D/\lambda }.  \label{dh1}
\end{equation}
The expression (\ref{dh1}) is derived without any spatial averaging over the
large- scale ( $\sim L$) structures, always present in a real flow where all
macroscopic characteristics, including the Lyapunov exponents $\lambda $,
are slightly modulated on the integral scale. Accounting for this spatial
variation gives an estimate following directly from (\ref{log1}): 
\begin{equation}
|\delta h_{r_{0}}|\sim L\sqrt{D/\lambda }\sqrt{\ln \left[ L/r_{0}\right] }.
\label{dh1L}
\end{equation}
The modified regime $A$, accounting for the large- scale averaging, will be
denoted hereafter by $A^{\prime }$.

\underline{B}. Pure inertial-convective range. Inertial-convective range is
realized at the scales $r\gg \eta \gg r_{0}$ where one cannot neglect
small-scale advection contribution . In the case of a general non-smooth
velocity one finds a strongly intermittent behavior, manifested in the
anomalous scaling of the scalar structure functions: 
\begin{equation}
S_{2n}(r)=\left\langle \left[ h({\bf r}_{1})-h({\bf r}_{2})\right]
^{2n}\right\rangle \sim L^{2n-\zeta _{2n}}r_{12}^{\zeta _{2n}},  \label{S2n}
\end{equation}
valid at the separations $r_{12}\ll L$. The fundamental origin of the
anomalous scaling, $\zeta _{2n}<n\zeta _{2}$, was discovered recently \cite
{95CFKLb,95GK,95SS}. It was understood that the anomalous exponents
originate from zero modes of the eddy-diffusivity operator: $\zeta _{2n}$
are universal numbers, solely defined by the velocity statistics and
independent on the properties of the pumping term. The exponents were
analytically calculated for the case of the velocity field, rapidly varying
in time, introduced by Kraichnan \cite{74Kra-a} in $1/d$-, $2-\zeta _{2}$- ,
and $\zeta _{2}$-expansions \cite{95CFKLb}, \cite{95GK}, \cite{95SS},
respectively. An instanton approach \cite{97Che} yields yet another large $n$
( $n\gg d$) asymptotic for $\zeta _{2n}\rightarrow \zeta _{\infty }(\zeta
_{2},d)$ when $n\rightarrow \infty $. The constant $\zeta _{\infty }(\zeta
_{2},d)$ was explicitly calculated. The saturation of exponents $\zeta _{n}$
was predicted also in \cite{97Yaka}). The instanton consideration, applied
to a general passive scalar problem, always results in the collapse of
exponents $\zeta _{2n}\rightarrow \zeta _{\infty }<\infty $ with the
asymptotic value $\zeta _{\infty }$ to be a complicated functional of the
velocity field statistics. It can be easily understood: the $2n$-th moment
of the scalar can be represented as a path integral over $2n$ fluid
particles. The dominant contribution into the high ( $2n$--th) order
structure function originates from the most probable $2n$- particle
trajectory. In the incompressible world it is impossible to avoid a
divergence of particles at least in one of the directions and it has been
shown in \cite{97Che} that contribution to the path integral from a single diverging
trajectory is sufficient to cause saturation of the exponents $\zeta _{n}$.
This effect has a very strong and important
influence on the parametric dependence of an effective 
front speed $u_{T}$ calculated below.

\underline{C}. Consecutive inertial-convective and dissipative-convective
ranges. The description of the scalar correlations in the
inertial-convective regime (ICR) does not deviate from the one considered
above for the case $B$. However, in the dissipative-convective interval
(DCR), $L\gg \eta \gg r\gg r_{0}$, the behavior of the scalar is very
different from that observed in the low-Reynolds- number Batchelor regime
when $L\approx \eta $. The crucial difference stems from the essential
non-Gaussianity and intermittency of the scalar field on the velocity
dissipation scale $\eta $ where the solutions in both intervals have to
match. In this case the scale $\eta $ is an integral scale for the DCR $%
r_{0}\ll r\ll \eta $ where powerful injection of all higher- order integrals
of motion ( $\int d{\bf r}$ $h^{n}$) takes place. The point is that a $2n-2$
moment defines a pumping for the next $2n$-th order one. Thus, considering a
generalization of the Kraichnan model for the case $C$( and neglecting the
correlations between the source and convective terms, which can slightly
renormalize some constants) one obtains the exact equations for the
multi-point correlation functions $F_{2n}({\bf r}_{1},\cdots ,{\bf r}%
_{2n})\equiv \left\langle h({\bf r}_{1})h({\bf r}_{2})...h({\bf r}%
_{2n})\right\rangle $,

\begin{equation}
\hat{L}F_{2n}=V^{11}(r_{12})F_{2n-2}({\bf r}_{3},\cdots ,{\bf r}_{2n})+perm.,
\label{LF}
\end{equation}
where $\hat{L}\equiv -{\cal K}^{\alpha \beta }({\bf r}_{ij})\nabla
_{i}^{\alpha }\nabla _{j}^{\beta }$ , $V^{\alpha \beta }(r)=K^{\alpha \beta
}(L)-K^{\alpha \beta }(r)$ and $K(r)\sim r^{2-\gamma }D$ at, $L>r>\eta $,
while $K(r)\sim r^{2}\eta ^{-\gamma }D$ at $\eta >r$. Operator $\hat{L}$ has
the $-\gamma $ dimensionality (it scales as $r^{-\gamma }$) in the
inertial-convective range while it is $O(r^{0})$ in the
dissipative-convective range. As a result, the $2n$-th moment $F_{2n}$,
explicitly depending on all $2n$-th points, is dominated by the forced
``logarithmic'' solution of (\ref{LF}) in the DCR. In the ICR, however, it
is a zero mode of the operator $\hat{L}$ which dominates the solution.
Therefore, the $2n$-th moment of the scalar difference, $S_{2n}(r)=\langle
\delta h_{r}^{2n}\rangle $ is estimated by 
\begin{equation}
S_{2n}(r)=a_{n}L^{2}\ln \left[ \frac{r}{r_{0}}\right] S_{2n-2},  \label{dhn}
\end{equation}
where $a_{n}$ are dimensionless $n$-dependent constants. On the another
hand, at the scale $\eta $ (\ref{dhn}) this expression should match the
anomalous structure functions (\ref{S2n}) from the upper
(inertial-convective) interval giving: 
\begin{equation}
S_{2n}(r)\sim \eta ^{\zeta _{2n}}L^{2n-\zeta _{2n}}\frac{\ln ^{n}\left[
r/r_{0}\right] ]}{\ln ^{n}\left[ \eta /r_{0}\right] }.  \label{dhn1}
\end{equation}

Let us proceed now with the second task and estimate the value of the
dissipative scale, $r_{0}$, required for calculation of the front velocity $%
u_{T}$. The point of a crucial importance is a necessity to distinguish
between the different moments of the front fluctuations at the dissipative
scale, $\delta h_{r_{0}}$. Usually one estimates the typical fluctuations as
a root-means-square value of the corresponding variable. This is fine in
case of ``normal scaling'' or if the small-scale intermittency is not too
strong (cases $A$ and $A^{\prime }$). However in the cases $B$ and $C$
intermittency in the convective intervals is extremely strong: estimates for
a typical front fluctuation based on various moments, $\delta
h_{r}^{(n)}\approx \left\langle \delta h_{r}^{n}\right\rangle ^{1/n}$ are
very different. As a result we get from (\ref{dh1},\ref{dh1L},\ref{S2n},\ref
{dhn1}) the following $n$-dependent estimate at the dissipative scale $r_{0}$
\begin{equation}
|\delta h_{r_{0}}^{(n)}|\sim L\left\{ 
\begin{array}{ll}
L\sqrt{D/\lambda }, & A, \\ 
\sqrt{D/\lambda }\sqrt{\ln \left[ L/r_{0}\right] }, & A^{\prime }, \\ 
\left[ r_{0}/L\right] ^{\zeta _{n}/n}, & B, \\ 
\left[ \eta /L\right] ^{\zeta _{n}/n}/\sqrt{\ln \left[ \eta /r_{0}\right] },
& C,
\end{array}
\right.   \label{dhn}
\end{equation}
for the cases $A,A^{\prime },B$ and $C$ respectively. On the other hand, the
value of $|\delta h_{r_{0}}|$ is defined by equilibration of the convective
and ''dissipative'' terms in (\ref{fpe1}) giving: 
\begin{equation}
|\delta h_{r_{d}}^{(n)}|\sim v_{F}\text{ }T,\hspace{0.1in}T\sim \left\{ 
\begin{array}{ll}
\lambda ^{-1}\ln \left[ L/r_{0}\right] , & A,A^{\prime }, \\ 
L/v_{rms}, & B,C.
\end{array}
\right.   \label{dhvf}
\end{equation}
It shows that the linear dimension of the flame brush, calculated on the
dissipative scale, is defined by the turbulent flame speed and the overall
time of Lagrangian evolution which is a typical time for Lagrangian
separation, initially equal to $r_{0}$, to reach the integral scale. In the $%
B,C$ regimes the overall time is $r_{0}$- independent (note that the time of
evolution from $r_{0}$ to $\eta $ is neglected in comparison with one
describing the inertial-convective stage of evolution, from $\eta $ to $L$).
Which number $n$ is essential to define $v_{F}$ via (\ref{dhvf}) is yet to
be discussed. One should add (\ref{first}) to (\ref{dhn},\ref{dhvf}).
Collecting all the relations in one table we get 
\end{multicols}
\begin{equation}
\begin{array}{ccccc}
& A & A^{\prime } & B & C \\ 
v_{F} & \frac{L\sqrt{D\lambda }}{\ln \left[ L\lambda /v_{0}\right] } & \frac{%
L\sqrt{D\lambda }}{\sqrt{\ln \left[ L\lambda /v_{0}\right] }} & 
v_{rms}\left[ \frac{v_{0}}{v_{rms}}\right] ^{\zeta _{n}\beta _{n}/n} & 
v_{rms}\left[ \frac{v_{0}}{v_{rms}}\right] ^{\zeta _{n}\beta _{n}/n}\ln
^{-1/2}\left[ \frac{\eta }{L}\left( \frac{v_{rms}}{v_{0}}\right) ^{\beta
_{n}}\right]  \\ 
r_{0} & \frac{v_{0}}{\lambda }\ln \left[ L\lambda /v_{0}\right]  & \frac{%
v_{0}}{\lambda }\ln \left[ L\lambda /v_{0}\right]  & L\left[ \frac{v_{0}}{%
v_{rms}}\right] ^{\beta _{n}} & L\left[ \frac{v_{0}}{v_{rms}}\right] ^{\beta
_{n}}
\end{array}
,\hspace{0.5cm}  \label{vfn}
\end{equation}
\begin{multicols}{2}
where, $\beta _{n}\equiv \left[ 1-\zeta _{1}+\zeta _{n}/n\right] ^{-1}$. As
was cited before, $\zeta _{2n}/[2n]\rightarrow 0$ and respectively, $\beta
_{n}\rightarrow 1/[1-\zeta _{1}]>0$, at $n\rightarrow \infty $. Since,
intermittency of the scalar is growing downscales from the scale of the
pumping, the higher ratio of the integral scale to the dissipative one (of
the turbulent velocity to the chemical one) the higher the moment-number $n$
entering the actual stationary velocity of the flame (which is defining the
actual width of the front). The highest value of $\delta h_{r_{0}}^{(n)}$ is
reached at $n\to \infty $. That is to say that the stationary speed of the
front is defined by such optimal and rare (instanton) configurations which
give the dominant contribution into the highest moments of the scalar field $%
h$. The optimal configurations define the maximal spreading of the flame
brush and control (mainly) the dissipation and the stationary velocity of
the brush finally. The necessity to account for the highest moments of the $%
\delta h_{r_{d}}$ has a simple Lagrangian explanation. The point of a
crucial importance is a failure of $2n$- particle description of the $2n$%
-order correlation function of the scalar at the scales smaller than the
dissipative one. This fact can be easily illustrated by a numerical
procedure attempting to describe the initially plane front in terms of
collection of $n$ Lagrangian particles \cite{AMY} each moving with velocity $%
u_{0}{\bf n}+{\bf v(x,t)}$ where ${\bf n}$ is a unit vector in the $z$%
-direction. The initial ( $t=0$) distance between these particles is $%
l_{0}\approx 1/n$. Very soon after beginning of the simulation the sharp
cusps started to form and one had to add more and more particles to preserve
continuity of the front. The development of these very sharp gradients,
where dissipation takes place, drives the necessary number of particle $%
n\rightarrow \infty $. We would like to reiterate a very important and
profound peculiarity of the situation: the expression (\ref{vfn}), taken at
the largest $n$, corresponds to a very rare optimal (instanton)
configuration, responsible for the highest moments of the scalar difference.
This means that to describe the small-scale ( $r\ll r_{0}$) dynamics of the
front height $h(t;{\bf r}$ $)$, one has to evaluate contributions from these
configurations. We postpone the more accurate investigation( which requires
some further development of the instanton technique of \cite{97Che}
accounting for the entire dynamics of the front surface) for a future
publication.

One of the principle results of this work, given by (\ref{vfn}) at $%
n\rightarrow \infty $, is that turbulent front speed strongly depends on the
ratio $\Gamma =r_{0}/\eta $, which is a novel dimensionless parameter of the
problem. Therefore in order to compare experimental data with theoretical
predictions we must first estimate $\Gamma $ and determine the front
propagation regime. Recent experimental studies of a passive front
propagating in turbulent media were performed in the flows generated by
vibrating grids, capillary waves, Taylor-Couette flow and Hell-Show cells.
All experiments (for the turbulent regime, $v_{rms}\gg v_{0}$) corresponded
to the case $C$. For example, in the vibrating grid experiments \cite{96SJR}
the values of $u_{rms}\approx 1~cm/sec$, $L\approx 1-10~cm,$ $Re\approx
10^{2}-10^{3}$ and $u_{0}/u_{rms}\approx 10^{-2}-10^{-3}$. (In the case of
the Kolmogorov turbulence one has an estimation $\eta \approx 10LRe^{-3/4},$
where $%
\mathop{\rm Re}%
$ is the Reynolds number and the factor $c\approx 10$ agrees with available
experimental data). For the experimental conditions of \cite{95RHR} one gets 
$\Gamma \approx 10^{-3}-10^{-4}$. The expression for the front velocity,
following from the Table (\ref{vfn}), evaluated at $r_{0}$ corresponding to $%
n\rightarrow \infty $ is:

\begin{equation}
v_{F}\sim \frac{u_{rms}}{\sqrt{\frac{2}{3}ln\left[ \eta /L\right] +ln\left[
U\right] }}_{U\rightarrow \infty }\rightarrow \frac{u_{rms}}{\sqrt{ln\left[
U\right] }},  \label{vF}
\end{equation}

\noindent where $U\equiv u_{rms}/u_{0}$. The rhs of (\ref{vF}) (similar
formula was derived in \cite{88Yak}, see also \cite{88Ker}, where the
dynamic renormalization group has been used for evaluation of the turbulent
flame speed $v_{F}$) agrees very well with experimental data on $v_{F}$ in a
variety of turbulent flows in a wide range of variation of the dimensionless
turbulent intensity $U\leq 20-500$. This universality can be readily
understood: in all the experimental situations the $O(ln\left[ \eta
/L\right] )$- contribution is not large and can be neglected in comparison
with the $O(ln\left[ U\right] )$-term. In a typical case $U\approx 20-100$
the transitional $u_{rms}$-based Reynolds number is: $Re_{c}\approx
10^{3}-10^{4}$, which is very high. This explains the relatively broad
applicability of the large $U$ asymptotic of (\ref{vF}) and therefore
challenge a higher $%
\mathop{\rm Re}%
$ experiment to test the general structure of (\ref{vF}).

We acknowledge support of a R.H.~Dicke fellowship (MC) and the ONR/URI grant
(VY).

\end{multicols}

\end{document}